\documentclass[a4paper]{jpconf}
\usepackage{txfonts}
\usepackage{epsfig}
\usepackage{graphicx}
\usepackage[square, sort&compress]{natbib}

\begin{document}

\title{Are short-term variations in solar oscillation frequencies the signature of a second solar dynamo?}

\author{Anne-Marie Broomhall$^1$, Stephen T. Fletcher$^2$,
David Salabert$^{3,4}$, Sarbani Basu$^5$, William J. Chaplin$^1$,
Yvonne Elsworth$^1$, Rafael A. Garc{\'{i}}a$^6$, Antonio
Jim\'enez$^{3}$, and Roger New$^2$}\address{$^1$ School of Physics
and Astronomy, University of Birmingham, Edgbaston, Birmingham B15
2TT, UK}
\address{$^2$ Faculty of Arts, Computing, Engineering and Sciences,
Sheffield Hallam University, Sheffield S1 1WB, UK}
\address{$^3$ Instituto de Astrof{\'{i}}sica de Canarias, E-38200 La
Laguna, Tenerife, Spain}
\address{$^4$ Departamento de Astrof{\'{i}}sica, Universidad de La
Laguna, E-38205 La Laguna, Tenerife, Spain} \address{$^5$ Yale
University, P.O. Box 208101, New Haven, CT 06520-8101, USA}
\address{$^6$ Laboratoire AIM, CEA/DSM-CNRS-Universit{\'e} Paris
Diderot, IRFU/SAp, Centre de Saclay, 91191 Gif-sur-Yvette, France}
\ead{amb@bison.ph.bham.ac.uk}

\begin{abstract}In addition to the well-known 11-year solar cycle,
the Sun's magnetic activity also shows significant variation on
shorter time scales, e.g. between one and two years. We observe a
quasi-biennial (2-year) signal in the solar p-mode oscillation
frequencies, which are sensitive probes of the solar interior. The
signal is visible in Sun-as-a-star data observed by different
instruments and here we describe the results obtained using BiSON,
GOLF, and VIRGO data. Our results imply that the 2-year signal is
susceptible to the influence of the main 11-year solar cycle.
However, the source of the signal appears to be separate from that
of the 11-year cycle. We speculate as to whether it might be the
signature of a second dynamo, located in the region of near-surface
rotational shear.
\end{abstract}

\section{Introduction}\label{section[introduction]}
The Sun is a variable star, whose magnetic activity shows systematic
variations. The most conspicuous of these variations is the 11-year
solar cycle \citep{Hathaway2010}. However, over the past twenty
years it has become apparent that significant (quasi-periodic)
variability is also seen on shorter timescales, between 1 and 2
years \citep[e.g.][]{Benevolenskaya1995,Mursula2003,
Valdes-Galicia2008}. Fletcher et al. \citep{Fletcher2010}
investigated the origins of this so-called ``mid-term'' periodicity
by looking at variations in the frequencies of solar oscillations.
Fletcher et al. used the Sun-as-a-Star observations made by the
Birmingham Solar Oscillations Network \citep[BiSON;][]{Elsworth1995,
Chaplin1996} and the Global Oscillations at Low Frequencies
\citep[GOLF;][]{Gabriel1995, Jimenez2003, Garcia2005} instrument
onboard the \emph{SOlar and Heliospheric Observatory} (\emph{SOHO})
spacecraft. In this paper we extend the work of Fletcher et al. by
examining data observed by the Variability of solar IRradiance and
Gravity Oscillations \citep[VIRGO;][]{Frohlich1995} instrument,
which is also onboard SOHO. VIRGO consists of three sun photometers
(SPMs), that observe at different wavelengths, namely the blue
channel (402\,nm), the green channel (500\,nm), and the red channel
(862\,nm). We have examined each set of VIRGO data individually and
find that the results are similar for each channel. Therefore, here
we concentrate on the results found using the blue VIRGO data.

The frequencies of solar p modes vary throughout the
solar cycle with the frequencies being at their largest when the
solar activity is at its maximum \citep[e.g.][]{Woodard1985,
Palle1989, Elsworth1990, Jimenez2003, Chaplin2007, Jimenez2007}. By
examining the changes in the observed p-mode frequencies throughout
the solar cycle, we can learn about solar-cycle-related processes
that occur beneath the Sun's surface.

We use oscillations data collected by making Sun-as-a-star
observations, which are sensitive to the p modes with the largest
horizontal scales (or the lowest angular degrees, $l$).
Consequently, the observed frequencies are of the truly global modes
of the Sun \citep[e.g.][and references therein]{JCD2002}. These
modes travel to the Sun's core but, because the sound speed inside
the Sun increases with depth, their dwell time at the surface is
longer than at the solar core. Consequently, p modes are most
sensitive to variations in regions of the interior that are close to
the surface and so are able to give a global picture of the
influence of near-surface activity.

Recently Garc{\'{i}}a et al. \citep{Garcia2010} observed signatures
of a stellar activity cycle in asteroseismic data obtained by the
Convection Rotation and Planetary Transits
\citep[CoRoT;][]{Baglin2006} space mission. With the prospect of
longer asteroseismic data sets ($\sim5\,\rm years$) becoming
available through, for example, Kepler \citep{Borucki2009}  there
will be opportunities to observe activity cycles in other stars.
These observations will provide constraints for models of stellar
dynamos under conditions different from those in the Sun.

\section{Uncovering the mid-term
periodicity}\label{section[results]}

The observations made by BiSON, GOLF and VIRGO were divided into
182.5-day-long independent subsets. BiSON has now been collecting
data for over 30 years. The quality of the early data, however, is
poor compared to more recent data because of poor time coverage.
Here, we have analyzed the mode frequencies observed by BiSON during
the last two solar cycles in their entirety i.e. from 1986 April 14
to 2009 October 7. GOLF and VIRGO have been collecting data since
1996 and so we have been able to analyze data covering almost the
entirety of solar cycle 23, i.e., from 1996 April 11 to 2009 April
7. After 1996 April 11, when all three sets of data were available,
we ensured that the start times of the subsets from each
observational program were the same.

Estimates of the mode frequencies were extracted from each subset by
fitting a modified Lorentzian model to the data using a standard
likelihood maximization method. Two different fitting codes have
been used to extract the mode frequencies, both giving the same
results. For clarity, we only show the results of one method, which
was applied in the manner described in \citep{Fletcher2009}. A
reference frequency set was determined by averaging the frequencies
in subsets covering the minimum activity epoch at the boundary
between cycle 22 and cycle 23. It should be noted that the main
results of this paper are insensitive to the exact choice of subsets
used to make the reference frequency set. Frequency shifts were then
defined as the differences between frequencies given in the
reference set and the frequencies of the corresponding modes
observed at different epochs \citep{Broomhall2009}.

For each subset in time, three weighted-average frequency shifts
were generated, where the weights were determined by the formal
errors on the fitted frequencies: first, a ``total'' average shift
was determined by averaging the individual shifts of the $l=0$, 1,
and 2 modes over fourteen overtones (covering a frequency range of
$1.88-3.71\,\rm mHz$); second, a ``low-frequency'' average shift was
computed by averaging over seven overtones whose frequencies ranged
from 1.88 to $2.77\,\rm mHz$; and third, a ``high-frequency''
average shift was calculated using seven overtones whose frequencies
ranged from 2.82 to $3.71\,\rm mHz$. The lower limit of this
frequency range (i.e., $1.88\,\rm mHz$) was determined by how low in
frequency it was possible to accurately fit the data before the
modes were no longer prominent above the background noise. However,
we note here that each of the fitted frequencies was checked for
accuracy and this resulted in many of the low-$n$ fitted frequencies
from the VIRGO data being discarded. The lower signal-to-noise of
the oscillations in the VIRGO data means that accurate fits to the
data are only possible above approximately $2.3\,\rm mHz$. The upper
limit on the frequency range (i.e., $3.71\,\rm mHz$) was determined
by how high in frequency the data could be fitted before errors on
the obtained frequencies became too large due to increasing line
widths causing modes to overlap in frequency.

\begin{figure*}
  \centering
  \includegraphics[width=0.36\textwidth, clip]{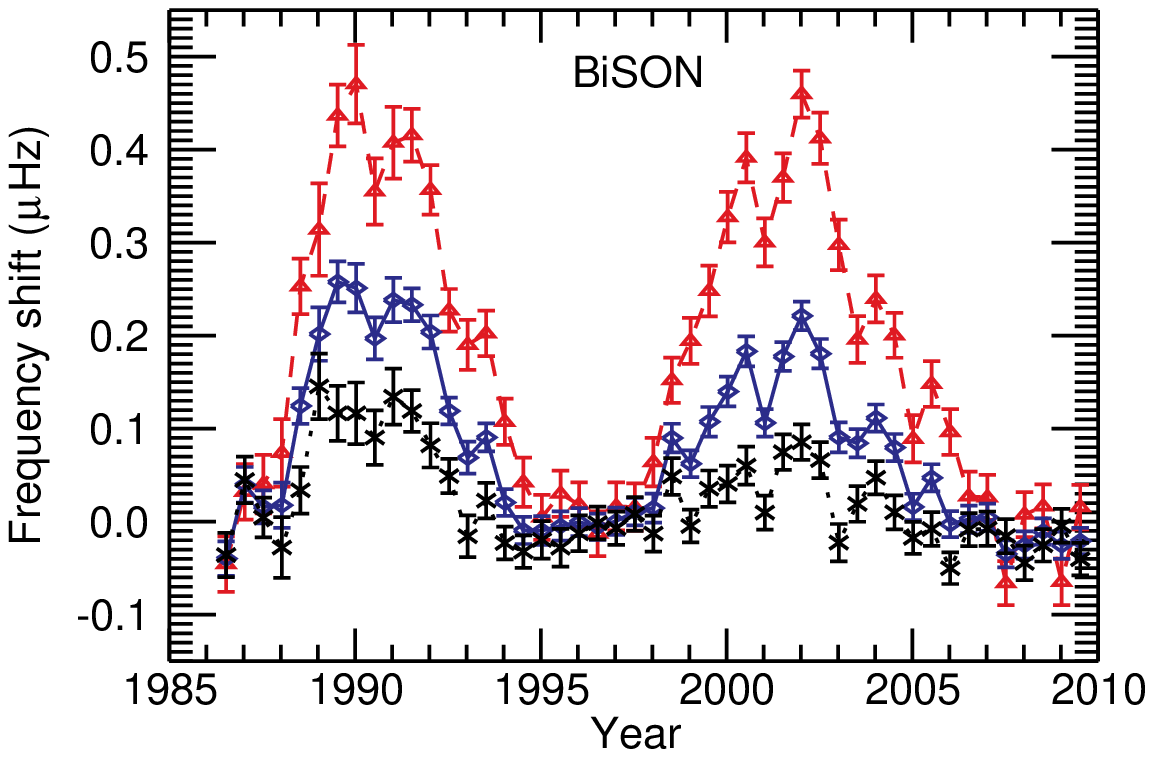}
    \includegraphics[width=0.36\textwidth,
    clip]{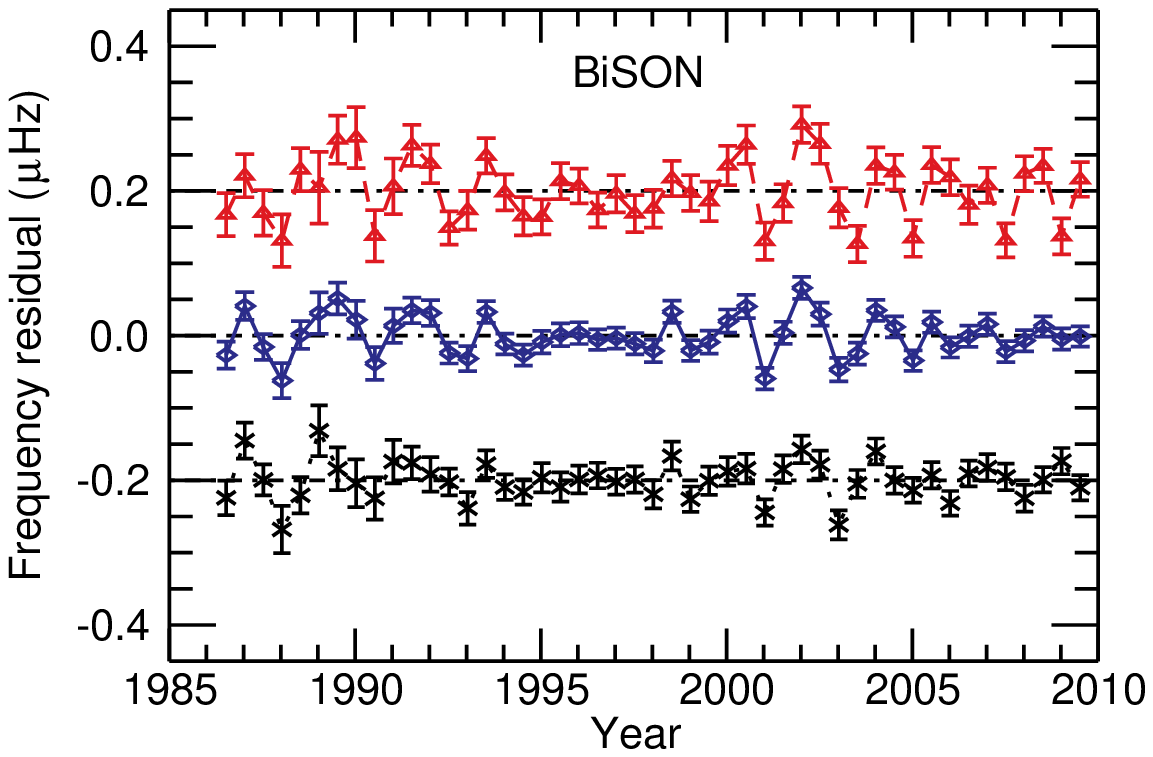}\vspace{0.2cm}\\
  \includegraphics[width=0.36\textwidth, clip]{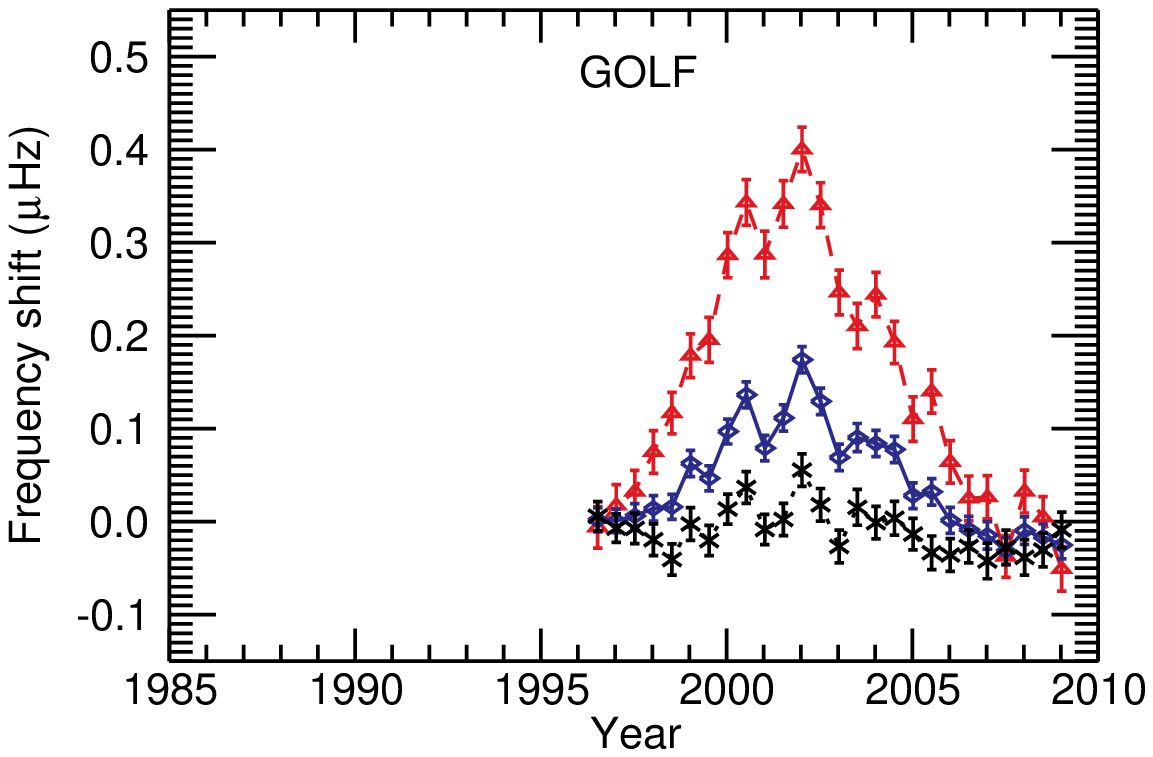}
  \includegraphics[width=0.36\textwidth,
  clip]{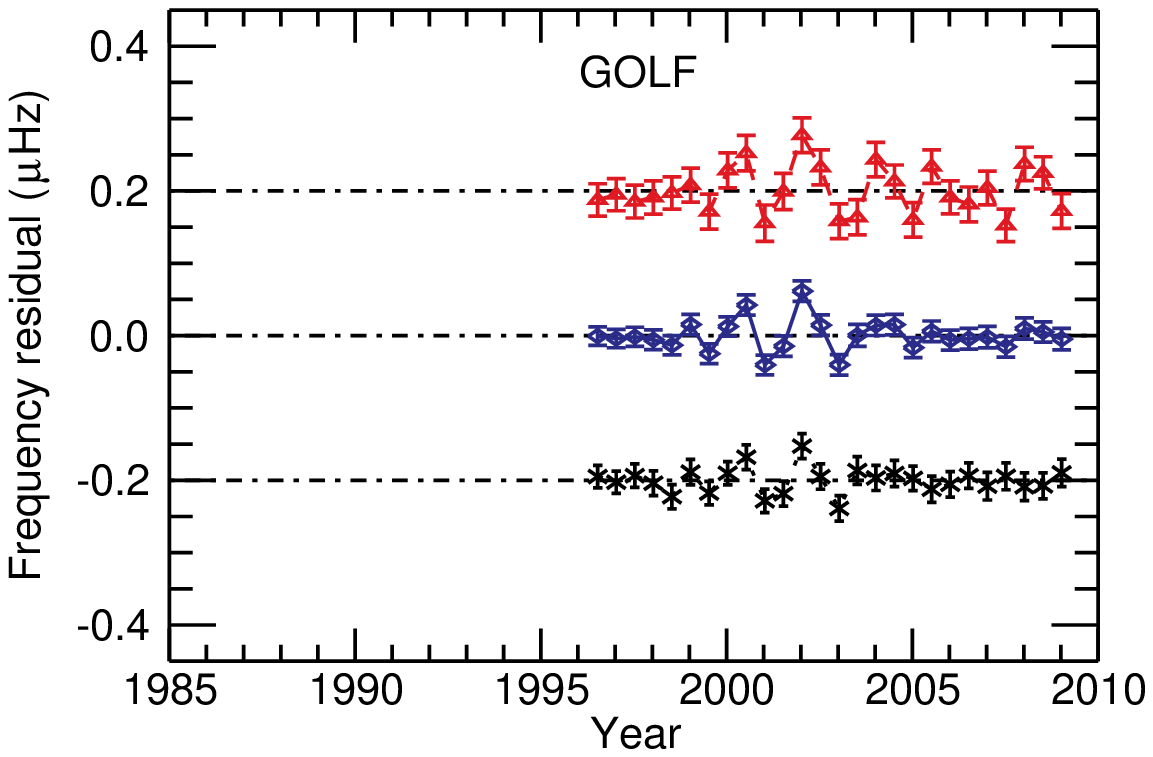}\vspace{0.2cm}\\
  \includegraphics[width=0.36\textwidth, clip]{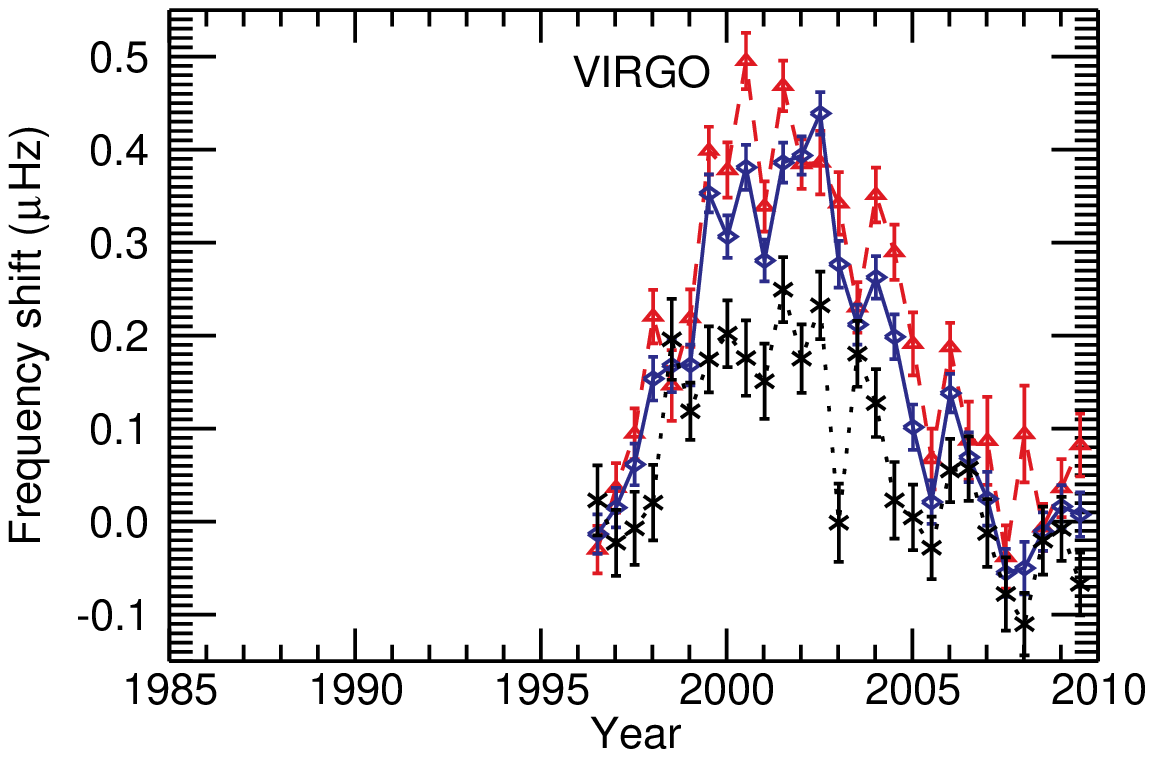}
  \includegraphics[width=0.36\textwidth, clip]{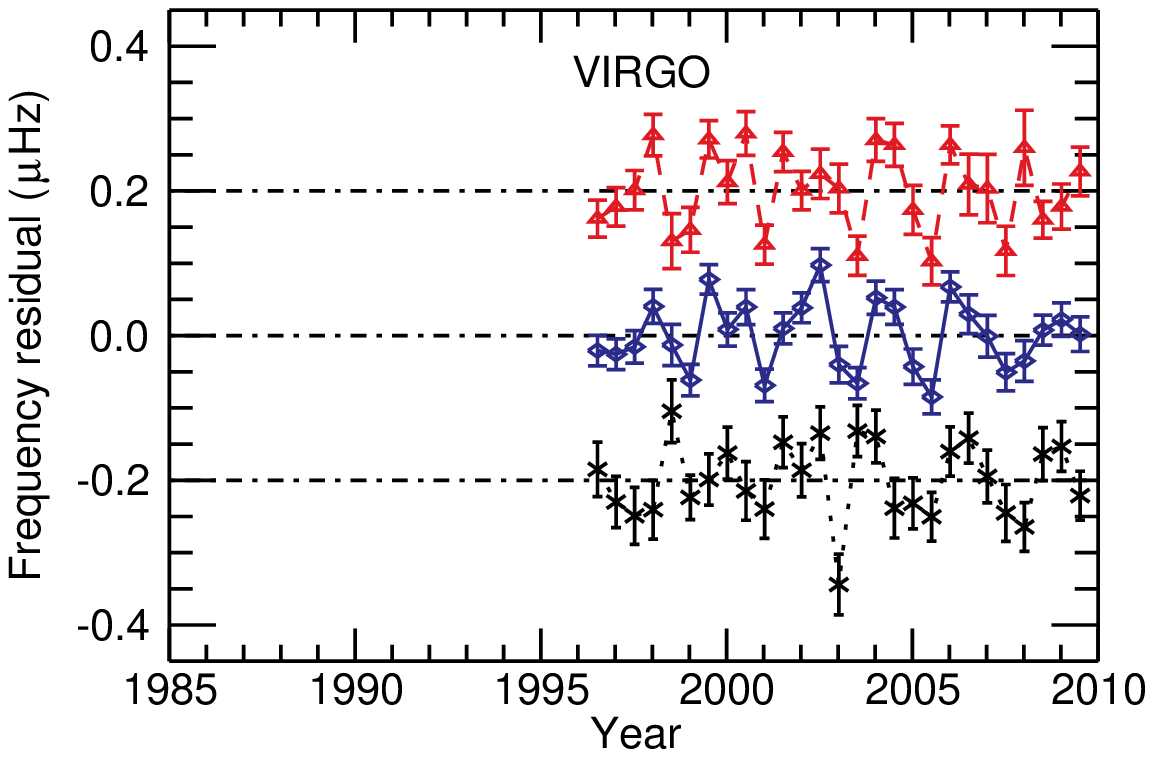}\\
  \caption{\small{Left column: average frequency shifts of ``Sun-as-a-star'' modes with
  frequencies between 1.88 and $3.71\,\rm mHz$ (total-frequency band, solid line, and diamond symbols); 1.88 and $2.77\,\rm
  mHz$ (low-frequency band, dotted line, and cross symbols); and 2.82 and $3.71\,\rm mHz$
  (high-frequency band, dashed line, and triangle symbols). Right column:
residuals left after dominant 11-year signal has been removed
(dotted and red dashed curves are displaced by $-0.2$ and $+0.2$,
respectively, for clarity).}}\label{figure[results]}
\end{figure*}

The left-hand panels of Figure \ref{figure[results]} show mean
frequency shifts of the p modes observed by BiSON, GOLF and blue
VIRGO, respectively \citep[also see][]{Broomhall2009, Salabert2009,
Fletcher2010}. The 11-year cycle is seen clearly and its signature
is most prominent in the higher-frequency modes. This is a telltale
indicator that the observed 11-year signal must be the result of
changes in acoustic properties in the few hundred kilometres just
beneath the visible surface of the Sun, a region that the
higher-frequency modes are much more sensitive to than their
lower-frequency counterparts because of differences in the upper
boundaries of the cavities in which the modes are trapped
\citep{Libbrecht1990, JCD1991}. Note that the difference between the
low- and high-frequency range shifts is less for the blue VIRGO
data, compared to the BiSON and GOLF data. This is because the
low-frequency range for the blue VIRGO data does not extend as low
in frequency as for the BiSON and GOLF data. Despite the low- and
high-frequency bands showing different sensitivities to the 11-year
cycle there is a significant correlation between the observed
frequency-shifts. The correlations between the low- and
high-frequency band shifts are 0.82 for the BiSON data, 0.67 for the
GOLF data, and 0.78 for the VIRGO data. The errors on the
correlations indicate that there is less than a 1\% chance that each
of these correlations would occur by chance. The signal is
reassuringly similar in the different data sets. The correlation
between the BiSON, GOLF, and blue VIRGO frequency shifts was found
to be highly significant in all three frequency bands with less than
a 1\% chance that these correlations would occur randomly.

In order to extract mid-term periodicities, we subtracted a smooth
trend from the average total shifts by applying a boxcar filter of
width 2.5 years. This removed the dominant 11-year signal of the
solar cycle. Although the width of this boxcar is only slightly
larger than the periodicity we are examining here, wider filters
produce similar results. The resulting residuals, which can be seen
in the right-hand panels of Figure \ref{figure[results]}, show a
periodicity on a timescale of about 2 years.

There is a significant correlation between the low- and
high-frequency band residuals for the BiSON (0.46), GOLF (0.55), and
blue VIRGO (0.76) data and there is less than a 1\% probability of
these correlations occurring by chance. The correlations between the
BiSON and GOLF residuals were found to be significant in all three
frequency bands with less than a 1\% probability of observing such
correlations by chance. The BiSON and blue VIRGO residuals are also
reasonably well correlated in all three frequency bands, with less
than a 2\% probability of observing the correlations by chance.
However, the GOLF and blue VIRGO residuals are less well correlated.

\begin{figure}
  \centering
  \includegraphics[width=0.36\textwidth, clip]{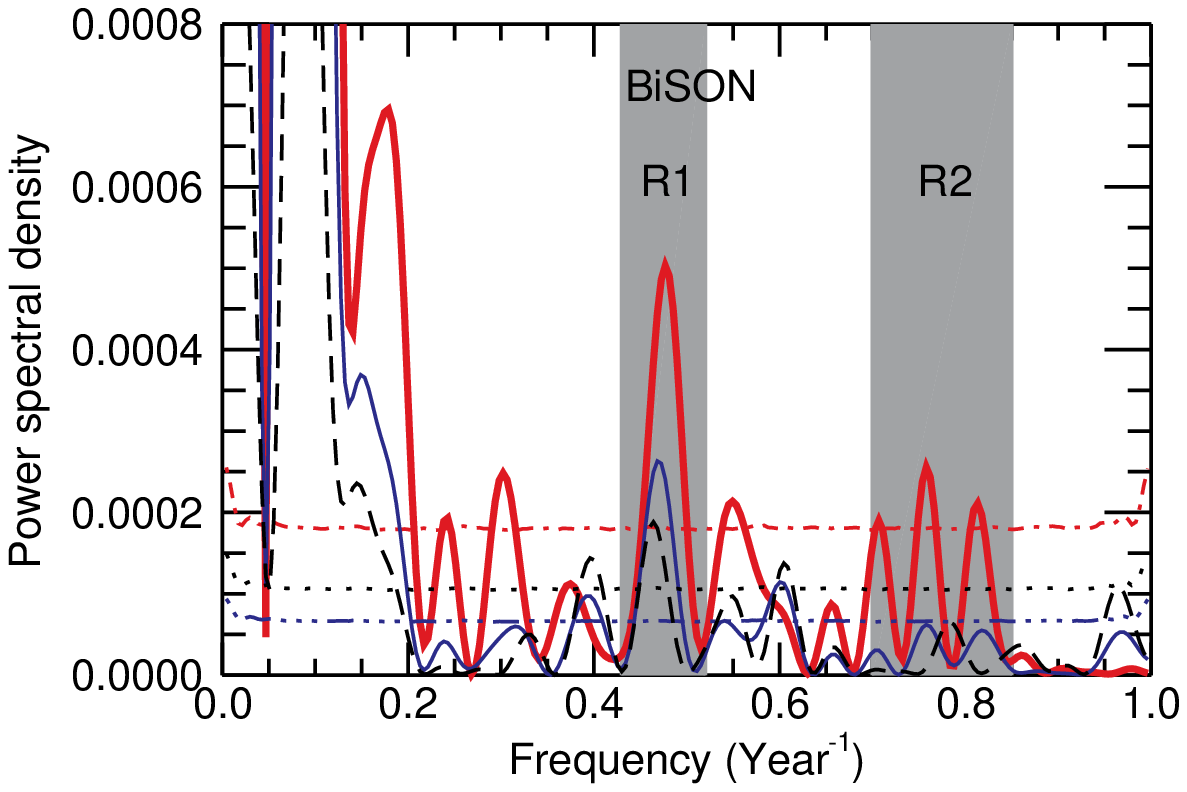}
  \includegraphics[width=0.36\textwidth, clip]{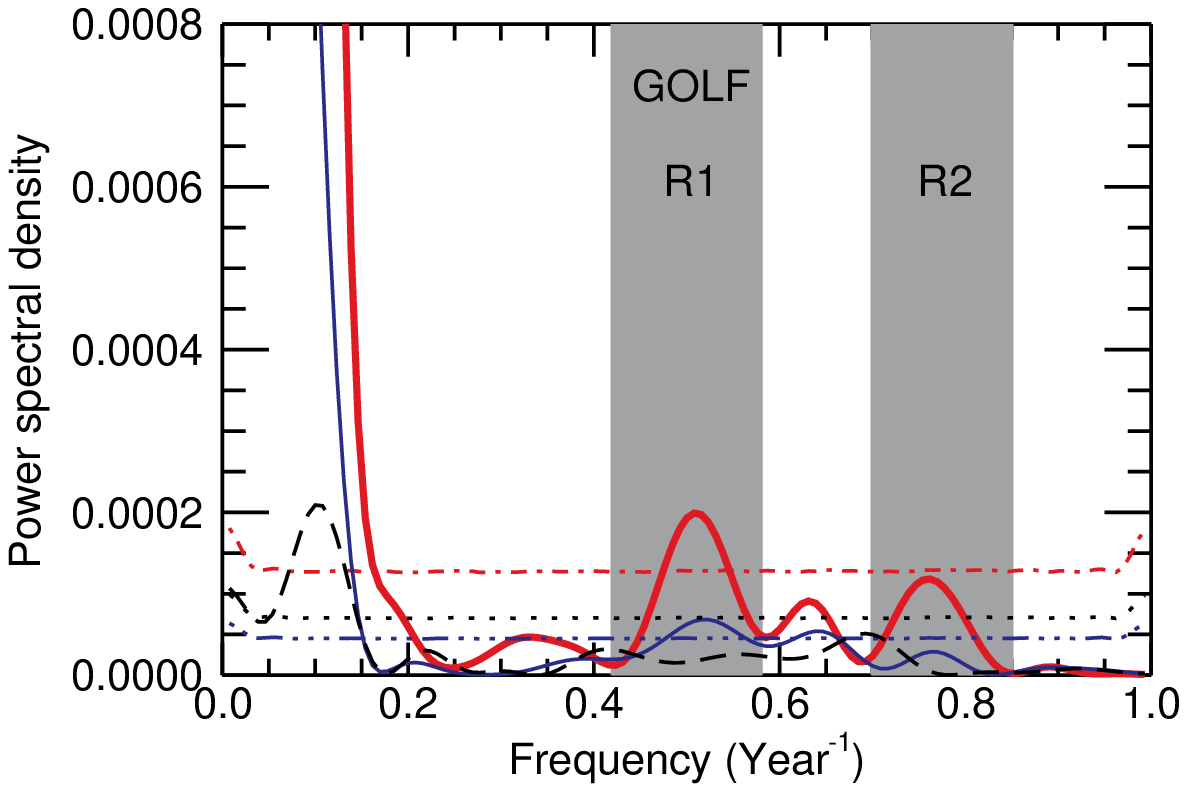}\\
  \includegraphics[width=0.36\textwidth, clip]{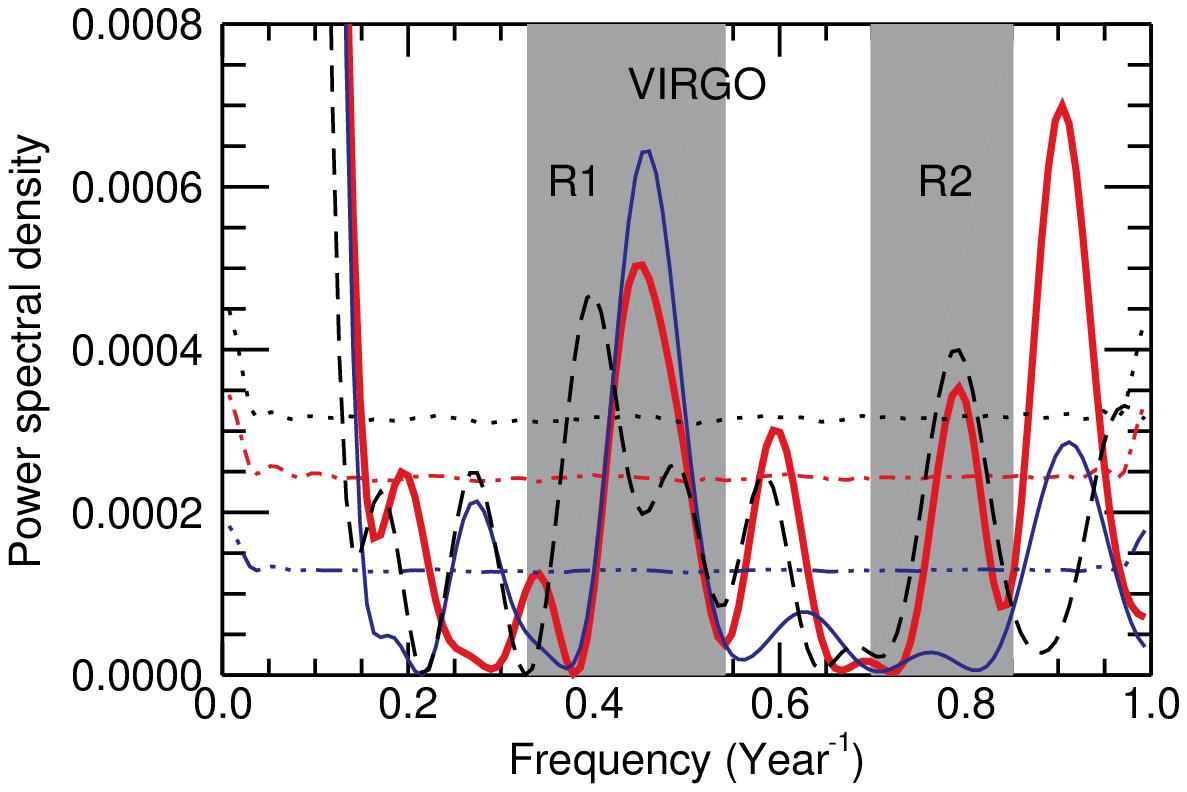}\\
  \includegraphics[width=0.48\textwidth, clip]{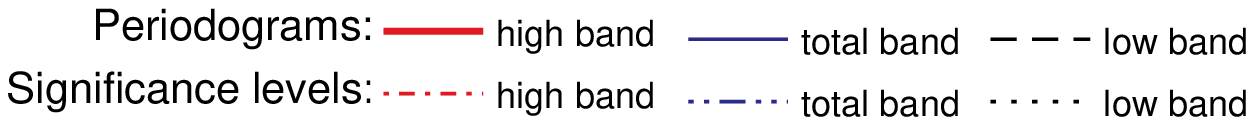}\\
  \caption{\small{Periodograms of the frequency shifts observed in the different frequency bands (see legend).
  The 1\% ``false alarm'' significance levels
  for the respective frequency ranges are also plotted. Shaded regions,
  denoted R1 and R2, are included in each panel to guide the eye towards the
significant regions of interest.}}\label{figure[periodogram]}
\end{figure}

The periodograms of the raw frequency shifts shown in the left-hand
panels of Figure \ref{figure[results]} were computed to assess the
significance of the 2-year signal. Figure \ref{figure[periodogram]}
shows the periodograms, oversampled by a factor of 10. Also plotted
in Figure \ref{figure[periodogram]} are the 1\% false alarm
significance levels \citep{Chaplin2002}, which were determined using
Monte Carlo simulations based on the size of the errors associated
with the raw frequency shifts (see Figure \ref{figure[results]}).
The large peak at $0.09\,\rm yr^{-1}$ is the signal from the 11-year
cycle. There are also large peaks at approximately $0.5\,\rm
yr^{-1}$ (indicated by the shaded regions denoted R1 in each panel
of Figure \ref{figure[periodogram]}). Statistical analysis of the
BiSON periodograms established that the apparent 2-year periodicity
was indeed significant, in the low-, total-, and high-frequency
bands, with a false alarm probability of 1\%. A peak at the same
frequency is also significant in the high- and total-frequency bands
in the GOLF and blue VIRGO data. Note that there is also a
significant peak at a slightly lower frequency in the low-frequency
band blue VIRGO data. The fact that the peaks in the GOLF and blue
VIRGO data are not as prominent as the equivalent peaks observed in
the BiSON data is expected because fewer GOLF and blue VIRGO data
are available, particularly during periods of high activity when the
2-year signal is most prominent.

The examination of the VIRGO data supports the theory that there is
a 2-year signal present in the frequency shifts. One possible
explanation for this signal is a dynamo action seated near the
bottom of the layer extending 5\% below the solar surface
\citep[see][for details]{Fletcher2010}. The amplitude envelope of
the 2-year signal observed in the BiSON and GOLF data appears to be
modulated by the 11-year cycle (see the right-hand panels of Figure
\ref{figure[results]}). This is particularly true for the
low-frequency band. Interestingly this does not appear to be the
case for the signal observed in the blue VIRGO data, which could be
because fewer very low-frequency modes were used to calculate the
blue VIRGO frequency shifts, thus indicating that the signal shows
some frequency dependence. Note that although asymmetries in the
Sun's magnetic field have been used to explain the 2-year signal
observed in other proxies of solar activity this would not explain
why the amplitude of the signal observed in the p-mode frequency
shifts is so similar in all frequency bands.

A prominent peak is observed at $\sim0.9\,\rm yr^{-1}$ in the blue
VIRGO high- and total-frequency bands. However, there is no signal
present at the same frequency in either the BiSON or GOLF data. This
peak could, therefore, be instrumental in origin.

\section{Evidence for a 1.3-year periodicity}
We also draw attention to a significant peak at a frequency of
approximately 0.8\,yr$^{-1}$ or a period of $\sim1.3\,\rm yr$
(indicated by the shaded regions denoted R2 in Figure
\ref{figure[periodogram]}). This peak is visible most strongly in
the blue VIRGO data but an excess of power is also visible in the
high-frequency range in the BiSON and GOLF data (although the GOLF
peak is only significant at a 2\% level). Notice that the 1.3-year
signal observed in the VIRGO data is significant in both the low-
and high-frequency bands but almost fully suppressed in the
total-frequency band. This is because the signal is out of phase in
the two different regions of the frequency-spectrum.

A $1.3\,\rm yr$ periodicity has been observed in other solar data.
For example, Howe et al. \citep{Howe2000} observed variations in the
rotation profile of the Sun, most predominately at low latitudes and
with a period of $1.3\,\rm yr$. However, the signal was found to be
intermittent and has not been observed since 2001 \citep{Howe2007}.
Jim{\'e}nez-Reyes et al. \citep{Jimenez2003} observed a $1.3\,\rm
yr$ modulation in the energy supply rate. Wang and Sheeley
\citep{Wang2003} observe a $1.3\,\rm yr$ quasi-periodicity in the
Sun's dipole magnetic moment and open magnetic flux. Wang and
Sheeley \citep{Wang2003} attribute this to the stochastic processes
of active region emergence and a decay time of about $1\,\rm yr$,
which is determined by differential rotation, meridional flow and
supergranule diffusion. The presence of excess power at this
frequency in all three sets of data means that this feature warrants
further investigation, as does the fact that the signal is out of
phase in the low- and high-frequency band blue VIRGO data, and so
the 1.3-year periodicity is the subject of ongoing work.

\ack This paper utilizes data collected by the Birmingham
Solar-Oscillations Network (BiSON), which is funded by the UK
Science Technology and Facilities Council (STFC). We thank the
members of the BiSON team, colleagues at our host institutes, and
all others, past and present, who have been associated with BiSON.
The GOLF instrument on board \emph{SOHO} is a cooperative effort of
many individuals, to whom we are indebted. \emph{SOHO} is a project
of international collaboration between ESA and NASA. S.T.F., A.M.B.,
W.J.C., Y.E., and R.N. acknowledge the financial support of STFC.
S.B. acknowledges NSF grants ATM-0348837 and ATM-0737770. D.S.
acknowledges the support of the grant PNAyA2007-62650 from the
Spanish National Research Plan. R.A.G. thanks the support of the
CNES/GOLF grant at the CEA/Saclay.

\bibliographystyle{iopart-num}
\bibliography{GONG_paper}

\end{document}